\documentclass[11pt]{revtex4}
\usepackage{graphicx}

\RequirePackage{amssymb}

\usepackage{amsmath}
\usepackage{latexsym}
\usepackage{amssymb}
\usepackage{graphicx}
\usepackage{dcolumn}
\usepackage{bm}

\input{epsf}
\begin{document}

\title{ Charge and Spin Currents in a Lower-Dimensional Supersymmetric Model in Presence of Vortices}
\author{Cristine N. Ferreira$^{a}$}
\email{crisnfer@iff.edu.br} 

\author{Jo\~ao P. S.  Alves e Silva $^{a}$}
\email{jpseixasesilva@hotmail.com}  

\author{Jos\'e A. Helay\"{e}l Neto$^{b}$}
\email{josehelayel@gmail.com}  

\author{N. Panza $^{c}$ }
\email{nelsonpanza@gmail.com }

\affiliation{$^{a}$Laborat\'orio de F\'{\i}sica e Suas Tecnologias,  \mbox{Instituto Federal de Educa\c{c}\~{a}o, Ci\^encias e Tecnologia Fluminense}, 
Rua Dr.\,Siqueira 273, Parque Dom Bosco, \mbox{Campos dos
Goytacazes, RJ,  Brazil, CEP 28030-130 }\\}

\affiliation{$^{b}$Centro Brasileiro de Pesquisas F\'{\i}sicas,  Rua Dr.\,Xavier Sigaud 150, Urca, Rio de Janeiro, RJ, Brazil, CEP 22290-180 \\}

\affiliation{$^{c}$Departamento de F\'{\i}sica, CEFET - RJ\\
\mbox{Avenida Maracan\~a, 229,  Maracan\~a, } \\
\mbox{ Rio de Janeiro, RJ,  Brazil, CEP  20271-110}}


\begin{abstract}
We investigate charge and spin currents that may appear in some materials, considering the possible couplings and the symmetries of a field-theoretical model presented here. We inspect these possible currents in (1+2) dimensions by adopting an ${\cal N}=2 - D=3$ supersymmetric framework with a Chern-Simons term and non-minimal couplings as well. We discuss a number of aspects in connection with a vortex configuration that is topologically viable. The new features of our investigation take into account the nature of both the U(1)-symmetry and supersymmetry (SUSY) breakings in presence of the vortex. We focus on aspects of the fermionic sector and their interactions with the currents. In connection with the spectrum of fermions, we also derive the mass gap in terms of the parameters of the the model. Another point we highlight is the role SUSY in our considerations. Once graphene and topological insulator materials are described by a Dirac-like equation, we assume, as a working hypothesis, that SUSY may also have some influence on the properties of these materials. Along this line, it is mandatory to connect the SUSY breakdown to this class of materials. We believe that our discussion could bring some elements for the understanding of this category of lower-dimensional systems. 
\end{abstract}


\maketitle


\newpage

\section{Introduction}

An important path to be followed for describing and understanding physical systems  
consists in analyzing their continuous global and local symmetries and eventual discrete symmetries as well.  
A crucial aspect associated to the presence of symmetries is, in most cases, the choice of a mechanism to break
them, in order to match the phenomenology, once not all symmetries are realized in its exact version. 
Symmetry breakdown is always present in phase transition phenomena, implying the existence  of stable topological defects \cite{Else:2018eas}. 
These topological defects may help us to understand the physical properties of the systems where they are present \cite{Liu:2018sww}. 
In analogy to what happens in a great deal of materials, an important point is the understanding of phase
transitions in the early Universe. The combination of a minimal coupling with gravity in an era dominated 
by post-inflationary kinetics, which usually appears in quintessential inflation scenarios, can lead to 
spontaneous symmetry breakings and eventually their restoration in the radiation age. Breaking these 
symmetries leads to the generation of short-lived topological defects that tend to produce 
gravitational waves until symmetry is restored \cite{Bettoni:2018pbl}.
The existence of such defects results from a symmetry breakdown triggered by the Higgs field in a primordial era, 
when the manifold's topology associated with this breaking is not simply-connected, so that an associated defect shows up. 
Such defects can be monopoles, cosmic strings or domain walls \cite{Sheng:2018jmy}. 
The interdisciplinary character of the subject allows us to borrow theoretical techniques from gravity and field theory 
to apply in condensed matter systems and, conversely, to provide examples of condensed matter to similar processes in the realm of gravity 
and field theory, but of much more difficult experimental observation. These topological defects can be experimentally implemented  
\cite{Matthews:1999zz,Schweikhard:2004zz,Eto:2011wp} and help us to obtain information about the systems in which they occur, as it is the case of the Bose-Einstein Condensate (BEC), where the defects can be formed by domain-wall annihilation \cite{Takeuchi:2012ee}. 
 In condensed matter systems, there are special materials whose dispersion relation for the energy is described
by a Dirac-type equation; that is the case for graphene and topological insulators. Topological insulator (TI) presents  an insulating
bulk gap generated by strong spin-orbit coupling and gapless metallic edge (2D
TI) or surface (3D TI) states. This behavior was observed in $Bi_xSb_{1-x}$ alloys,  in $Bi_2Se_3$, $Bi_2Te_3$
and $Sb_2Te_3$ crystals \cite{Hsieh2008} .  In such materials, there is an analogy with gravitational systems. When these 
materials are deformed into a cone shape, a current appears on their walls related to the deficit angle. This angle is similar 
to the one that occurs in cosmic-string topological defects when coupled to gravity \cite{Liu:2019xvw,Martin-Ruiz:2018wum,Fonseca:2011fe}. 
Even in the gravitational case, it is necessary to study the potentials involved and the breakings to get a better understanding of the system. 
In this paper, we study the symmetry and supersymmetry breakings that can generate stable solutions to give us a better 
understanding of what is needed to generate charge and spin currents, considering a (1+2)-dimensional model. The topological defect - in our case a vortex configuration - is contemplated in an ${\cal N}=2 - D=3$ supersymmetric description with a Chern-Simons and non-minimal couplings. In the literature, we can refer to a work that studies the effects of ${\cal N}=( 2,2) $-supersymmetry and classifies possible SUSY-preserving boundary conditions on charged matter fields around the vortex defects \cite{Hosomichi:2017dbc}. In our  approach, we adopt the 
fact that the ${\cal N}=1 $-supersymmetric theory in (1+3) dimensions with a vortex is preserved  for  ${\cal N}=2 $-supersymmetry  in (1+2) dimensions.
 Supersymmetry has raised a great deal of interest as it is considered a fundamental symmetry where fermionic excitations are naturally 
accommodated and is often used to stabilize theoretical models, such as the Standard Model of Particle Physics. Just as Dirac equation 
fits well to approach the physics of low-dimensional systems and may describe materials such as graphene, it can be thought 
that supersymmetry can also bring novelties in the interpretation of these types of materials \cite{Hosomichi:2017dbc}.
In this sense, a supersymmetric framework may become an efficient form to understand the correct couplings, 
and its breaking can be important to realize some effects in the material \cite{Abreu:2010yv}.  
p-form potentials appear in many supersymmetric models. A 2-form field
is referred to as the Kalb-Ramond field (KR) \cite{Barone:2010zi,KR,Lund}. In 4D, this
2-form gauge potential is on-shell equivalent to a real scalar field and may become relevant
in situations where the mediating particles are massless scalars\cite{Francia05}. 
The equivalence between a scalar field and a 2-form gauge potential is important
to understand the spontaneous symmetry breaking induced by the KR field in a Goldstone-like model. 
KR fields are also important in the study of the  vortex superfluids\cite{Davis88,Davis89,Ferreira08}.  
They have already been studied  in the physics of topological insulators \cite{Cho:2010rk}.
Considering that supersymmetry breaking can have an effect on the formation of topological 
defects in the universe, in this work, we shall pursue an investigation of the inverse problem, namely, 
which role supersymmetry may play in the appearance of currents on the surface of materials that present deformations.
 Our present contribution is outlined as follows: In the Section II, we present the bosonic and fermionic  
field components, the topological defect - vortex configurations and implications of the potential  configurations
to  the currents of the system. In this Section, we also inspect the conditions for vortex formation in terms of the parameters of the 
model and analyze the breaking of both the internal symmetry and supersymmetry. We choose those solutions that exhibit new effects.
In Section III, we focus on a graphene-like system in the presence of a topological vortex. 
We study the specific {\it ansatz} that yields the correct description and a comparative analysis of the parameter space of the model 
is carried out by assessing the evolution of the mass gap. The study of couplings, which allows us to better understand the currents that appear in these types of systems, are discussed in Section IV. In this Section, we also consider the emergence of spin and charge currents and their effects. Finally, in the Concluding Comments, we consider the highlight and summarize the main features of our study and comment on new prospects and incoming activities.

\section{Discussion of the framework with Hall Effect}

We adopt the point of view of introducing supersymmetry since we claim that the fermion-boson symmetry may help us to understand the relationship between the potential minima that describe topological defects and their associated currents. The model we inspect presents two Dirac fields with global symmetries, necessary to describe the systems analyzed in this contribution. For dimensions (1+3), the model relies on the Kalb-Ramond (2-form) gauge field with a non-minimal coupling. In this Section, we focus on the implications of the dimensional reduction to a planar world, discuss the appearance of vortex configuration and consider the issue of supersymmetry breaking.

\subsection{Analysis of fermionic and bosonic  fields from the  ${\cal N}=1- D=4$ dimensional reduction}
Now, let us consider the notation we adopt and the (1+2)-dimensional field content from the (1+3)-dimensional splitting of fields. 
The (1+3)-D superfield action with the Kalb-Ramond coupling 
\cite{Christiansen:1998xf} is given by
\begin{equation}
S_{4D} = \int \it d^4x \, d^2\theta \Big\{ -{1 \over 8} {\cal W}^a{\cal W}_a + d^2\bar \theta \Big[ -{1 \over 2} {\cal G}^2 + {1 \over 2} \Delta_{CS} {\cal V}{\cal G} + {1\over 16} \bar \Phi e^{2 q {\cal V}} \Phi e^{4 g {\cal G}} \Big] \Big\}
\end{equation}
where $ \Delta_{CS} $ is the Chern-Simons constant coupling, $q$ stands for the charge and $g$ is the coupling constant of the non-minimal coupling. There appear three superfields in this formulation: the chiral scalar superfield, $ \Phi (\phi , \chi , S)$, where $\phi $ is a complex scalar field, $\chi $ is a complex fermion and S is a complex auxiliary field:

\begin{eqnarray}
\Phi \, (\phi , \chi , S) = e^{-i \theta \sigma^{\hat \mu} \bar{ \theta} \partial_{\hat \mu}} \Big(\phi(x) + \theta^a \chi_a(x) + \theta^2S(x) \Big), & \bar D_{\dot{a}} \Phi =0 
\end{eqnarray}
with $D_a$ and $\bar D_{\dot{a}}$ are the supersymmetry covariant derivatives,
\begin{eqnarray}
D_a &= & \partial_a - i \sigma^{\hat \mu}_{\,\, a \dot{a}} \bar{\theta}^{\dot a}\partial_{\hat \mu} \\
\bar{ D}_{\dot{a}}  &=&- \partial_{\dot{a}} + i \theta^{a} \sigma^{ \hat \mu}_{\,\, a\dot{a}} \partial_{\hat \mu}
\end{eqnarray}
The field strength superfield ${\cal W}_a $, is written as 
\begin{equation}
{\cal W} ^a =-{1 \over 4} \bar {\cal D}^2{\cal D}^a {\cal V} 
\end{equation}
where the gauge superfield that contains the electromagnetic field, $A_{\hat \mu}$, is ${\cal V}(A_{\hat \mu}, \lambda , D) $; in the Wess-Zumino (WZ) gauge, its component-field expansion is as follows:
\begin{eqnarray}
{\cal V} (A_{\hat \mu}, \lambda , D) =  \theta \sigma^{ \hat \mu} \bar \theta A_{\hat \mu} + \theta^2 \bar{\theta \lambda} + \bar{\hat \theta}^ 2 \theta \lambda + \theta^2 \bar{\theta}^2 D (x)
\end{eqnarray}
According to our conventions, the index  $\hat \mu = 0, 1, 2, 3 $  refers to four-dimensional space-time. This superfield also accommodates  the gaugino field, $ \lambda$,  and a real auxiliary field, $D $.

The  term  responsible for the BF-mixing displays the coupling between the electromagnetic superfield, ${\cal V} $, and the Kalb-Ramond field strength superfield, ${\cal G}( M, \xi, \tilde G_{\mu}) $, where $M$ is the real scalar field and $ \xi$ is the fermionic component field. In this superfield, no  auxiliary field is present. The Kalb-Ramond field strength superfield is $ {\cal G} = {i \over 8} ( D^a \Sigma_a - \bar D_{\dot a} \bar \Sigma^{\dot a} ) $, where  $ \Sigma_\alpha $ is a chiral spinor superfield:

\begin{eqnarray}
\Sigma_a &=&   \psi_a(x) +\theta^b \Omega_{b a}(x) + \theta^2\Big[\xi_a(x) + i \sigma^{\hat \mu}_{\,\, a\dot{a}} \partial_{\hat \mu} \bar{\psi}^{\dot a} (x) \Big] - i \theta \sigma^{\hat \mu} \bar \theta \partial_{\hat \mu} \psi_a(x) \\
& = & i \theta \sigma^{\hat \mu }\bar{\theta} \theta^b \partial_{\hat \mu }\Omega_{b a}(x) - {1\over 4} \theta^2 \bar{\theta}^2 \Box \psi_a(x)
\end{eqnarray}
$\Omega_{b a} = \epsilon_{ba} \rho(x) + \sigma^{\hat \mu}_{\,\, b a}{\cal B}_{\hat \mu \hat \nu}(x)$, with $\rho(x)$ and ${\cal B}_{\hat \mu \hat \nu} (x)$ are complex fields given by
\begin{eqnarray}
\rho (x) &=& P(x) + i M(x),\\
{\cal B}_{\hat \mu \hat \nu}(x) &=& {1 \over 4} \Big[ B_{\hat \mu \hat \nu}(x) - i \tilde B_{\hat \mu \hat \nu} (x)\Big]
\end{eqnarray}
with $\tilde B_{\hat \mu \hat \nu} (x) = {1 \over 2} \epsilon_{\hat \mu \hat  \nu \hat \alpha\hat \beta} B^{\hat \alpha \hat \beta}(x) $ and ${\cal B}_{\hat \mu \hat \nu} = i {\cal B}_{\hat \mu \hat \nu} $, then the component-field expansion for ${\cal G}$ is given by

\begin{eqnarray}
{\cal G} ( M, \xi, \tilde G_{\hat \mu}) &=& -{1\over 2} M + {i \over 4} \theta^a \xi_a + {i \over 2} \theta^a \sigma^ {\hat \mu}_{\,\, a \dot{a}} \bar{\theta}^{\dot{a}} \tilde G_{\hat \mu} \\
& = & {1 \over 8} \theta^a \sigma^ {\hat \mu}_{\,\, a \dot{a}}  \bar{\theta}^2 \partial_{\hat \mu} \bar{\xi}^{\dot{a}} - {1 \over 8} \theta^2 \sigma^{\hat \mu}_{a \dot a} \bar{\theta}^{\dot a} \partial_{\hat \mu} \xi^a - {1\over 8} \theta^2 \bar{\theta}^2 \Box M
\end{eqnarray}

In this formulation, there is the dual of the Kalb-Ramond field strength, $ \tilde G_{\hat \mu}, $ given by

\begin{eqnarray}
\tilde G_{\hat \mu} &=& {1 \over 3 !} \epsilon_{ \hat \mu \hat \nu \hat \alpha \hat \beta} G^{ \hat \nu \hat \alpha \hat \beta } \label{dualkr}\\
G_{\hat \nu \hat \alpha  \hat \rho}& = & \partial_{\hat \nu } B_{\hat \alpha  \hat \rho} + \partial_{\hat \alpha} B_{\hat \nu \hat \rho} +  \partial_{\hat \rho} B_{\hat \nu \hat \alpha}\label{kr}
\end{eqnarray}
where  $G_{\hat \mu \hat \nu} = \partial_{\hat \mu} B_{\hat \nu} - \partial_{\hat \nu} B_{\hat \mu} $ with $\hat \mu , \hat \nu= 0, 1, 2,3 $. Every time we refer to (1+2) space-time indices, we adopt the notation of $\mu$ without hat:  $ \mu= 0, 1, 2 $; $\hat \mu = \mu, 3$. 
We are interested in considering a graphene-like structure. These systems are described by the Dirac equation in (1+2) dimensions; so, it is important to proceed to a dimensional reduction. Accordingly, in three-dimensional space, we shall carry out the following identifications $N= A^3$, $B^\mu = B^{3 \mu}$, $G^{\mu \nu} = \partial^{\mu}B^{\nu} - \partial^{\nu}B^{\mu} $ and $B^{\mu \nu} =\epsilon^{\mu \nu \rho} Z_{\rho} $, $ \partial_{\mu}Z^{\mu} = - \tilde G^{3} $, where $ \tilde G^3$ is related to $\partial_\mu Z^\mu $.

Then, the reduction of the fields to (1+2) dimensions, considering $\partial_3(fields) =0$, is given by
\begin{center}
\begin{tabular}{|p{3.5cm}||p{7cm}|}\hline %
\multicolumn{2}{|c|}{Table I: The splitting of the bosonic fields } \\
\hline
(1+3) dimensions & (1+2) dimensions \\
\hline \hline
&\\
$- {1 \over 6} G_{\hat \mu \hat \nu \hat \rho}G^{\hat \mu \hat \nu \hat \rho} \, $ & $ -{1 \over 2} G_{ \mu \nu} G^{ \mu \nu}  + Z^2  $  \\
& \\  \cline{1-2}
&\\
$-{1 \over 4} F_{\hat \mu \hat \nu} F^{\hat \mu \hat \nu}$&$-{1 \over 4} F_{\mu  \nu} F^{ \mu  \nu}  +  {1 \over 2} \, \partial_\mu N \, \partial^\mu N $    \\  
&\\ \cline{1-2}
&\\
$ \Delta_{CS} \epsilon_{\hat \mu \hat \nu \hat \rho  \hat \lambda } A^{\hat \mu} \partial^{\hat \nu}B^{\hat \rho \hat \lambda}  $& $2 \Delta_{CS} \epsilon_{\mu  \nu \rho  } A^{\mu} \partial^{\nu}B^{ \rho} +  2 \Delta_{CS} \, N \,Z $ \\ 
&\\ \cline{1-2}
&\\
$\nabla_{\hat \mu} \phi^*\nabla^{\hat \mu} \phi  e^{-2 g M}$ &$\nabla_{\mu} \phi^*\nabla^{\mu} \phi e^{-2 g M} +( h \, N - g\, Z)^2 \, \phi^* \phi  e^{-2 g M}$\\ 
& \\\cline{1-2}
\end{tabular}
\end{center}
We cast, in  Table I, the splitting of the bosonic fields upon dimensional reduction. The Kalb-Ramond field in (1+3) dimensions presents three different couplings. One of these is the bilinear of the Kalb-Ramond field strength, with $G_{\hat \mu \hat \nu \hat \rho} $ given by (\ref{kr}); the other is related to the Kalb-Ramond-Chern-Simons term that contains the coupling between the Kalb-Ramond and the gauge potentials. This is responsible for the Chern-Simons term in (1+2) dimensions, with two vector gauge fields. In graphene-like systems, this field can help us to understand the coupling to an external magnetic field. The last relevant term  appears inside of the covariant derivative, given by $
\nabla_{\hat \mu} = \partial_{\hat \mu} + iq A_{\hat \mu} + ig \tilde G_{\hat \mu}  $
with $\tilde G_{\hat \mu} $ defined in (\ref{dualkr}). 
In (1+2) dimensions, we adopt that $\hat \mu = ( \mu, x_3 ) $ with $ \mu =0,1, 2 $, the $ \tilde G_{\mu } ={1 \over 2} \epsilon_{ \mu \nu \alpha } G^{  \nu \alpha}$, then the covariant derivative is given by 
\begin{eqnarray}
 \nabla_{\mu} &=& \partial_{ \mu} + iq A_{ \mu} + ig \tilde G_{ \mu} \label{covderivative} \\
 \nabla_3 &= & i(q \,N - g\, Z)
 \end{eqnarray}
Here, we consider only the $\Psi$ and $\Lambda$  four-dimensional fermionic sector and we  take into account the splitting of the $\Gamma^{\hat \mu}$ Dirac matrices in four space-time dimensions given by
\begin{equation}
\Gamma^{\mu} = \left(\begin{array}{cc}
\gamma^\mu & 0 \\
0 & - \gamma^\mu
\end{array}\right) \, \, \, \,  \Gamma^{3}  =\left(\begin{array}{cc}
0 &\, \, \, \, \, \, i \\
i & \, \, \, \,  \, \, 0
\end{array}\right)\, \, \, \,  \Gamma_{5}= i \Gamma^0 \Gamma^1 \Gamma^2\Gamma^3  = \left(\begin{array}{cc}
0 &\, \, \, \, \, \, \gamma_3 \\
- \gamma_3& \, \, \, \,  \, \, 0
\end{array}\right)
\end{equation}
with $\gamma_3= I_2  $
We re-write the fermionic fields as four-component Majorana spinors:
\begin{eqnarray} 
\Psi \rightarrow  \xi , \zeta  \,\,\,\,\, \Lambda \rightarrow \lambda, \eta 
\end{eqnarray}
which gives rise to the following Dirac spinors
\begin{eqnarray} 
\Psi_{\pm} \rightarrow  \xi \pm i \zeta  \,\,\,\,\, \Lambda_{\pm} \rightarrow \lambda \pm i \eta \ \label{splitfermion}
\end{eqnarray}
\begin{center}
\begin{tabular}{|p{5cm}||p{10cm}|}\hline %
\multicolumn{2}{|c|}{Table II: The fermionic splitting for the $\Psi $- and $\Lambda$-fields } \\
\hline
(1+3) dimensions & (1+2) dimensions  \\
\hline \hline
& \\
$ {i \over 4} \bar \Psi \Gamma^{\hat \mu } \partial_{\hat \mu}\Psi$ & ${i \over 4}( \bar \Psi_+  \gamma^{ \mu } \partial_{ \mu}\Psi_+ + \bar \Psi_-  \gamma^{ \mu } \partial_{ \mu}\Psi_-)  $  \\ 
&\\ \cline{1-2}
&\\
$ {i \over 2} \bar \Lambda  \Gamma^{\hat \mu } \partial_{\hat \mu}\Lambda$&$ {i \over 2}( \bar \Lambda_+  \gamma^{ \mu } \partial_{ \mu}\Lambda_+ + \bar \Lambda_-  \gamma^{ \mu } \partial_{ \mu}\Lambda_-) $    \\ 
& \\ \cline{1-2}
&\\
$-{g^2 \over 4 q}\bar \Psi \Gamma_5 \Gamma^{\hat \mu}{\cal J}_{\hat \mu}\Psi e^{-2g M } $ &$-{g^2 \over 4 q}((\bar \Psi_+\gamma^{\mu}{\cal J}_{\mu}\Psi_+ - \bar \Psi_- \gamma^{\mu}{\cal J}_{ \mu}\Psi_-)e^{-2g M } 
 + {\cal O} ( N,Z)$\\ 
&\\ \cline{1-2}
&\\
$i(  \Delta_{CS}  + g q \phi^* \phi e^{-2g M }   )\,  \bar \Lambda \Gamma_5 \Psi $ &$i \, (  \Delta_{CS}  + g q \phi^* \phi e^{-2g M }   )\,(\bar \Lambda_- \Psi_+ - \bar \Lambda_+ \Psi_- )  $\\ 
&\\ \cline{1-2}
\end{tabular}
\end{center}
The  current ${\cal J}_\mu$ turns out given by 
\begin{eqnarray}
{\cal  J}_{\mu} = - {i q \over 2}(\phi^* \nabla_{\mu} \phi - \phi \nabla_{\mu}\phi^*) \label{current}
\end{eqnarray}
we can use the charge conjugation property, $\Psi_+ = (\Psi_-)^C$ and $\Psi_-= (\Psi_+)^C$, and write as \cite{Christiansen:1998xf}
\begin{eqnarray}
{\cal O}( N, Z) &=& -{g^2 \over 4}( \bar \Psi_+ \Psi_+ -\bar \Psi_- \Psi_- )(q N - g Z) e^{-2gM}
 \end{eqnarray}
 In  Table II,  we consider the reduction of the fermionic fields from (1+3) dimensions. The interesting terms here are the four-component Majorana spinors that can be organized in terms of two-components spinors in (1+2) dimensions. Both these fields have a global symmetry and can describe properties of the graphene. The spin-orbit term is given by the coupling of the fermions with the bosonic current ${\cal J}_\mu $. The important sector of the SUSY transformations  reads as follows

\begin{eqnarray}
\delta N &=& -{1 \over 2} ( \bar \varepsilon_+ \Lambda_+ + \bar \varepsilon_- \Lambda_- )\\
\delta \Lambda_\pm &=& (2 D + i \gamma^\mu \partial_\mu N) \varepsilon_\pm \pm \gamma_\mu \tilde F^\mu \varepsilon_\pm)\\
\delta D &=& -{i \over 2}(\bar \varepsilon_+ \gamma^\mu \partial_\mu \Lambda_+ + \bar \varepsilon_- \gamma^\mu \partial_\mu \Lambda_- )\\
\delta \tilde F^{\mu} &=& -{1 \over 2} ( \bar \varepsilon_+ \epsilon^{\mu \nu \lambda} \gamma_\lambda \partial_\nu \Lambda_+ -  \bar \varepsilon_+ \epsilon^{\mu \nu \lambda} \gamma_\lambda \partial_\nu \Lambda_-)
\end{eqnarray}
the tensor multiplet component fields transform according to

\begin{eqnarray}
\delta M & = & {i \over 4}( \bar \varepsilon_+ \Psi_- - \bar \varepsilon_- \Psi_+) \\
\delta \Psi_\pm &=& \pm 2 (\gamma^\mu \partial_\mu M + i Z ) \varepsilon_\mp - 2 i \gamma_\mu \tilde G^ \mu \varepsilon_\mp \\
\delta Z& =& {1 \over 4} (\bar \varepsilon_+ \gamma^\mu \partial^\mu \Psi_- - \bar \varepsilon_- \gamma^\mu \partial_\mu \Psi_+ ) \\
\delta \tilde G^\mu &=& - {i \over 4} (\bar \varepsilon_- \epsilon^{\mu \nu \lambda} \gamma_\lambda \partial_\nu \Psi_+ + \bar \varepsilon_+ \epsilon^{\mu \nu \lambda} \gamma_\lambda \partial_\nu \Psi_-)
\end{eqnarray}

With this prescription, the fields $\varphi, M,N$ present in $m_1$ and $\alpha$  interact with the fermions through the non-minimum coupling and they are related with the vortex configuration. The parameters that appear in $m_2$ are also associated to the non-minimum coupling and they shall be related with the graphene-like configuration as we wish to show in the next Section.

In Figure 1, we implement numerically the potential of the theory considering a possibility where it can recover the usual vortex configuration. In this case, the expected values of $<M>$ and $<N>$ both equal to zero can be considered. If it is so, the only contribution to the expectation value of the vortex field is the constant $\kappa_1$ from the Fayet-Iliopoulos term.

\subsection{Numerical analysis of the solutions of the potential generated by the vortex configuration in (1+2) dimensions}

In this Section, we analyze  the  full vortex configuration in (1+2) dimensions considering the convenient vortex gauge equations.  We consider the vortex configuration in the ${\cal N} =2, D=3$ model that results from the dimensional reduction of the four-dimensional CSKR model \cite
{Christiansen:1998xf,Ferreira:2009xm}.  It is necessary to study the stability of the configuration and the relation between all parameters in our approach. This model descends from the ${ \cal N} =1, D=4$ action that describes QED in the supersymmetric version coupled to the Kalb-Ramond field in a non-minimal way.  In the ${\cal N} =2, D=3$ model \cite{Becker:1995sp}, we write down the sector of gauge fields and the bosonic action in terms of components as:
\begin{eqnarray}
S_{gauge} &=&\int d^{3}x\{-{\frac{1}{4}}F_{\mu \nu }F^{\mu \nu
}+\,2\,\Delta_{CS}\,\varepsilon ^{\mu \nu \alpha }A_{\mu }\partial _{\nu }B_{\alpha } 
-{\frac{1}{2}}G_{\mu \nu }G^{\mu \nu }\},  \label{CS}
\end{eqnarray}
where the index $\mu =0,1,2$, with $F_{\mu \nu }=\partial _{\mu }A_{\nu
}-\partial _{\nu }A_{\mu }$ being the electromagnetic field-strength. $
B_{\mu }$ is a vector  with a corresponding field-strength $G_{\mu
\nu }=\partial _{\mu }B_{\nu }-\partial _{\nu }B_{\mu }$.  
The origin of the coupling constant $\Delta_{CS}$ is given by  the Kalb-Ramond and gauge superfield coupling in (1+3) dimensions. 

The part of the ${\cal N} =2 - D=3$ action involving the scalars is written as follows:
\begin{equation}
S_{scalar}=\int d^{3}x \Big[(\nabla _{\mu }\phi)^* \nabla ^{\mu }\phi \, e^{-2 g M} +{1 \over 2 } \partial_{\mu} N \partial^{\mu} N + (1- g^2 \phi \phi^* e^{-2gM}) \partial_{\mu} M \partial^{\mu} M \Big]
\label{scalar}
\end{equation}
 The covariant derivative, $\nabla _{\mu },$ is given by
(\ref{covderivative}) and the dual fields, $F_{\mu }={\frac{1}{2}}\epsilon _{\mu \nu \kappa }F^{\nu \kappa }$ and 
$G_{\mu }=\frac{1}{2}\epsilon _{\mu \nu \kappa }G^{\nu \kappa }$.

 The theory contains two auxiliary fields, one of them is the vector $ \cal {V} $, that we refer to as "D", and the other is the Z-field, which does not exhibit any dynamics in the action of the model. To find the potential in this case, we include two Fayet-Iliopoulos term in the action.
The action that gives us the potential of the vortex configuration, $S_{D,Z}$,  is given by
\begin{eqnarray}
S_{D,Z}& =& \int d^{3}x \{\,2q D |\phi |^{2}e^{-2g M}+2D ^{2} +\kappa_1 D - 4\Delta_{CS}M D \nonumber \\
&+&  Z^2 + 2 \Delta_{CS} N Z  -(q N- g Z)^2 \phi \phi^* e^{-2gM} + \kappa_2 Z\}. \label{potaction}
\end{eqnarray}
Now, we can search for field configurations arising from the potential. The equation of motion for the auxiliary field follows below:
\begin{eqnarray}
D &=&\Delta_{CS}M-{\frac{q}{2}}|\phi |^{2}e^{-2g M}-{\frac{\kappa_1 }{4}.} \label{Delta} \\
Z &=& - {(\Delta_{CS} + q g \phi \phi^* e^{-2gM})N + {\kappa_2 \over2} \over (1 - g^2 \phi \phi^* e^{-2gM})}
\end{eqnarray}
then, the scalar potential with the $\partial_\mu Z^\mu  = Z$ and interaction with the N-field takes the form:
\begin{eqnarray}
U &=&{\frac{q^{2}}{2}}\Big( |\phi |^{2}e^{-2g M}-  {2\Delta_{CS} \over q}M+ {\kappa_1 \over 2q}\Big)^2\nonumber \\& +  &{[(\Delta_{CS}
+  g q\phi \phi^* e^{-2g M} )N+ {\kappa_2 \over2}]^2 \over (1 - g^2 \phi \phi^* e^{-2gM})} +q^2N^2 \phi \phi^* e^{-2gM}\label{Pot1}
\end{eqnarray}
Considering the case with N given by

\begin{equation}
N   = -{\kappa_2 \over 2} {  (\Delta_{CS} +q \, g \, \phi \phi^* e^{-2 g M}) \over \Delta_{CS}^2+   2\, q\, g \Delta_{CS} \, \phi \phi^* e^{-2 g M} + q^2 \phi \phi^* e^{-2 g M}} \label{N}
\end{equation}
which also satisfies the extreme values for N, we can get several important cases that may be applied to realistic situations.

The new features of the present model are the non-minimal coupling and the Chern-Simons term. The main point is that we can apply it to describe a graphene-type material when it is subject to impurities or deformations. Supersymmetry may be an ingredient to render the model stable.
 \begin{figure}[htb]
\begin{center}
\includegraphics[width=6cm, height=6cm]{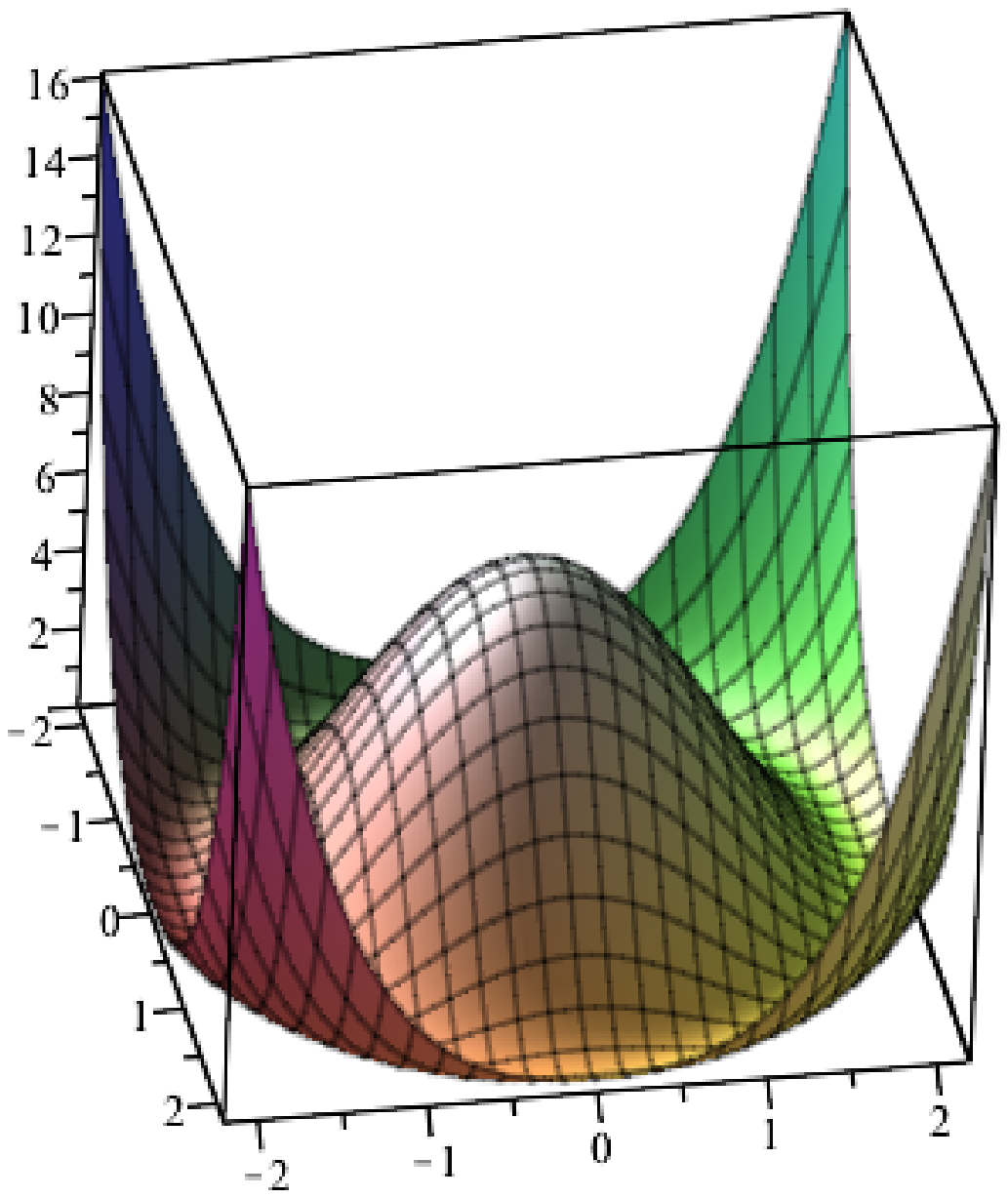}
\includegraphics[width=6cm, height=5cm]{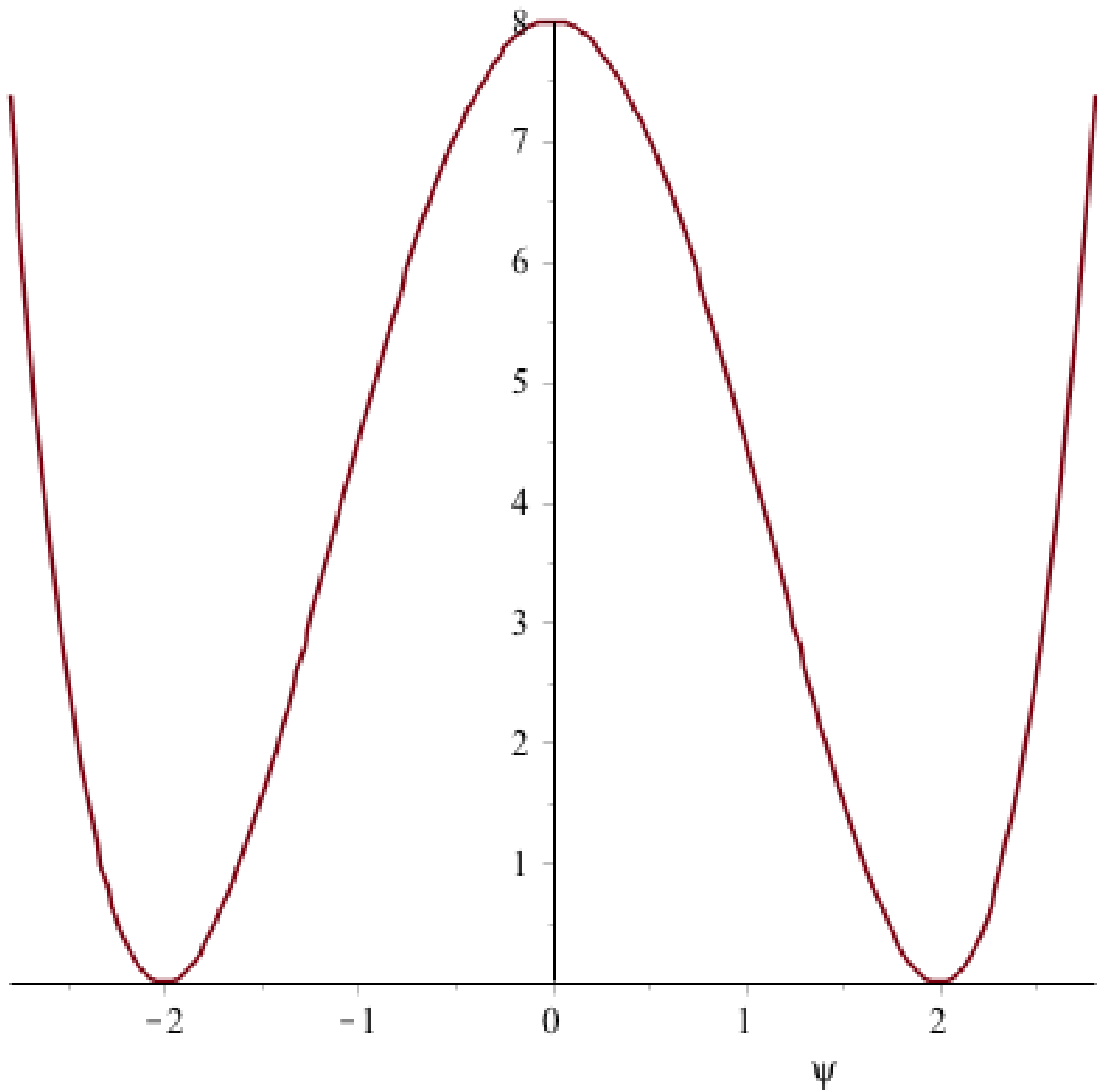}
\caption{ Vortex potential with the vacuum value  is given by $<\phi> = 2$  with the maximum in $<\phi>=0 $ considering $<M> =0$,  $\kappa_1 = -8$, $\kappa_2=0 $,  $\Delta_{CS} =1 $, $q=1$, $g=1$ that give us $<N>= 0$. }
\end{center}
 \label{Figure1}
 \end{figure}
In Figure 1, we implement numerically the potential of the theory considering a possibility where it can recover the usual vortex configuration. In this case the expected values of the $<M>$ and $<N>$ equal to zero can be considered. In this case the only contribution to the expected value of the vortex particles is the constant $\kappa_1$ from the term of Fayet-Iliopoulos $D$. We can see that the maximum point occurs at point  $\phi_1 = 0$ and $\phi_2 = 0$, if we write $\phi \phi * = \phi_1^2 + \phi_2^2 $ and there is a circle of minima. In the core of the vortex, SUSY is broken; this has been analyzed by the points that makes zero the potential. If the potential in the extremes is vanishing, then the supersymmetry is not broken; but, if it is non-zero, spontaneous breaking takes place.
 \begin{figure}[htb]
\begin{center}
\includegraphics[width=6cm, height=6cm]{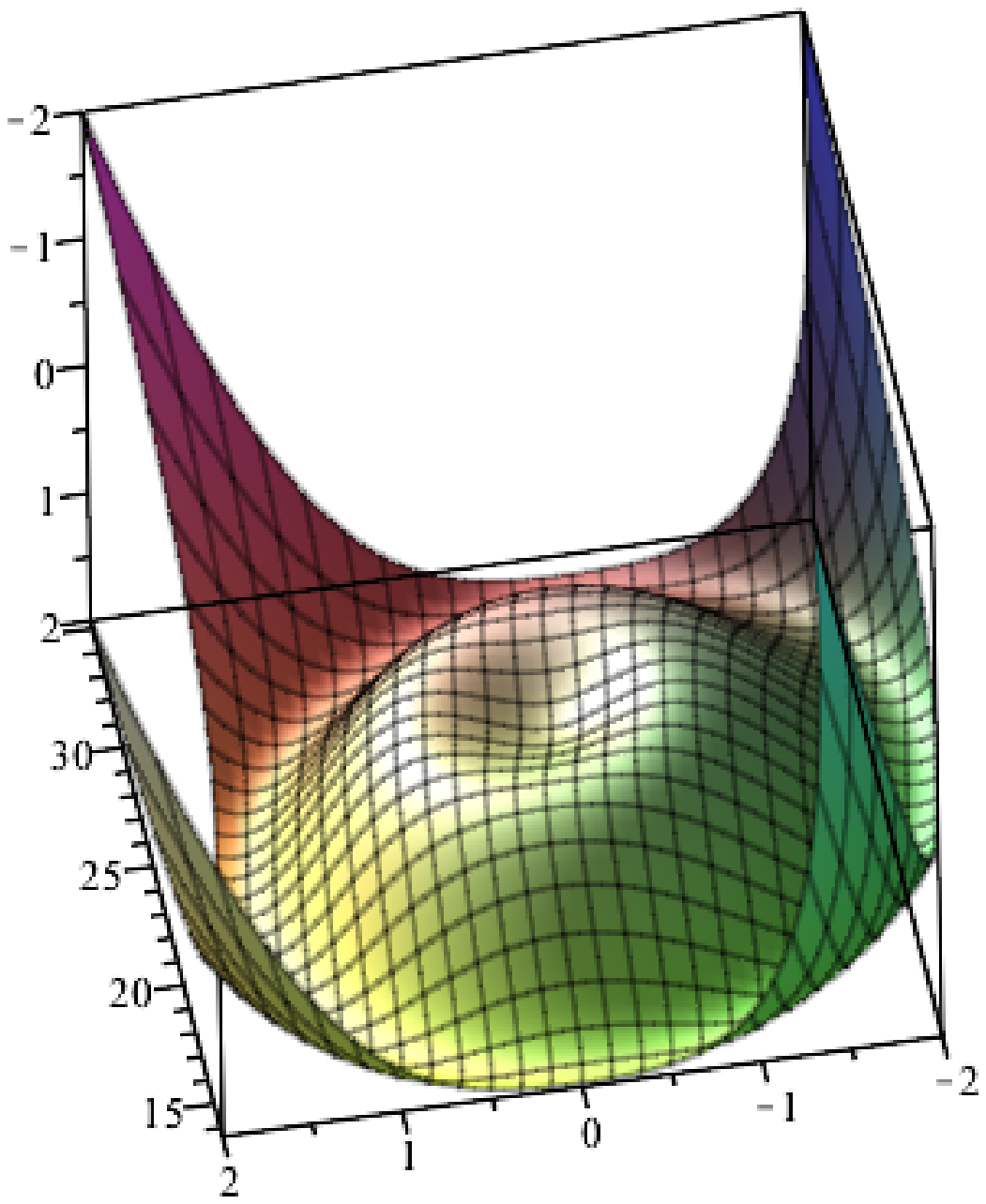}
\includegraphics[width=6cm, height=6cm]{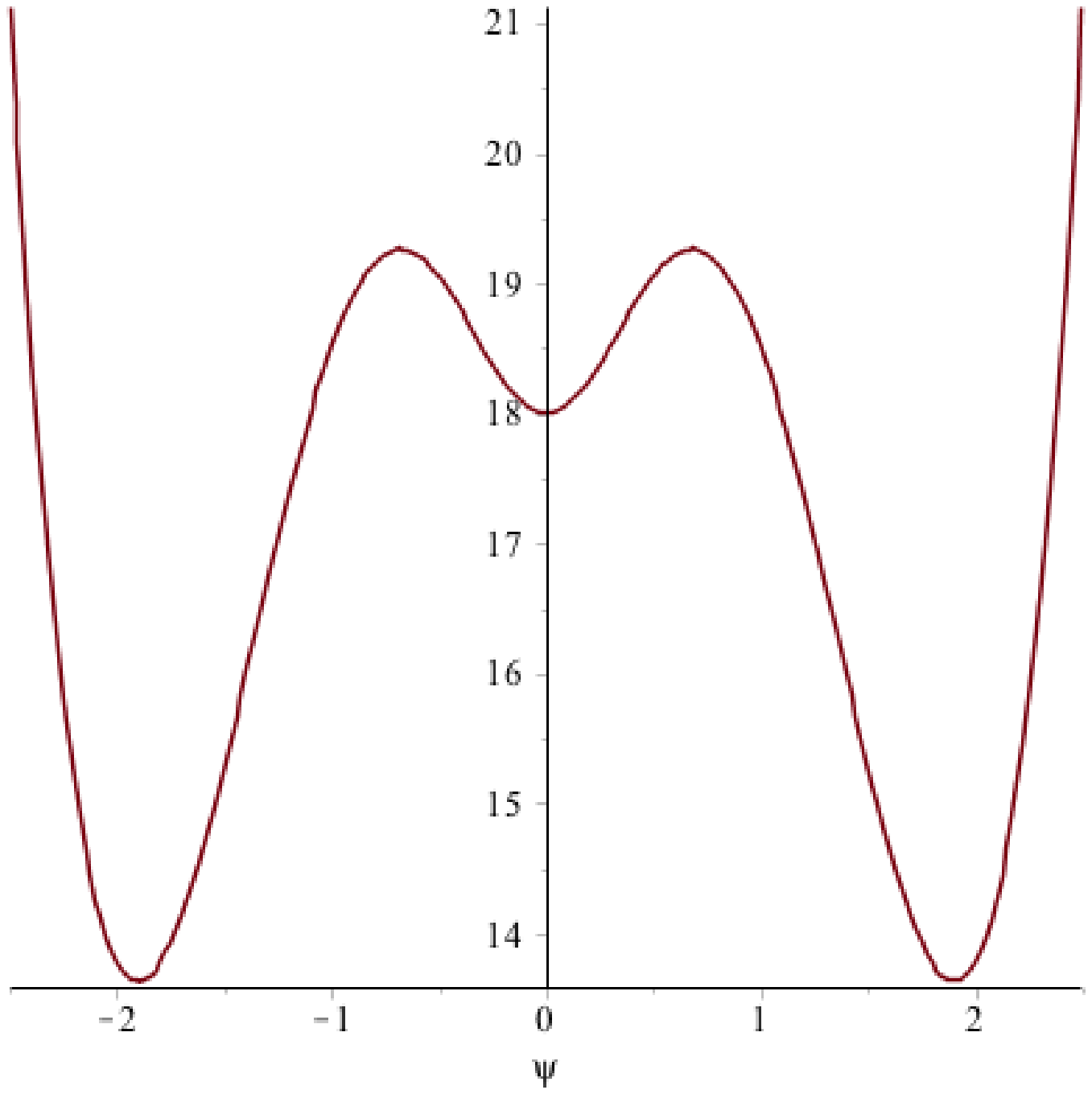}
\caption{
In this plot, we exhibit the situation in which the vacuum $ <M>=1.2$ and the parameters  $\kappa_1 = -8$, $\kappa_2=-6.5$,  $\Delta_{CS}=1 $, $q=1$ and $g=- 0.2$.   }
\end{center}
 \label{Figure2}
 \end{figure}
 We see, from Figure 2, that, in this case, we have two types of minima; in the core, we have a single minimum,  $<\phi>  = 0$. This result gives some light on certain systems obtained by shrinking graphene, where there is a migration of charges to the top of a cone-shaped deformed graphene film. In this case, the charges are polarized at the top, making a kind of Hall effect. The other type of minimum gives us the usual minimum circle. In the discussion Section, we shall analyze this behavior in detail.
 In this case, both extremes break supersymmetry, so we can say that it is the SUSY breaking that generates this effect from a theoretical point of view.
 \begin{figure}[htb]
\begin{center}
\includegraphics[width=6cm, height=6cm]{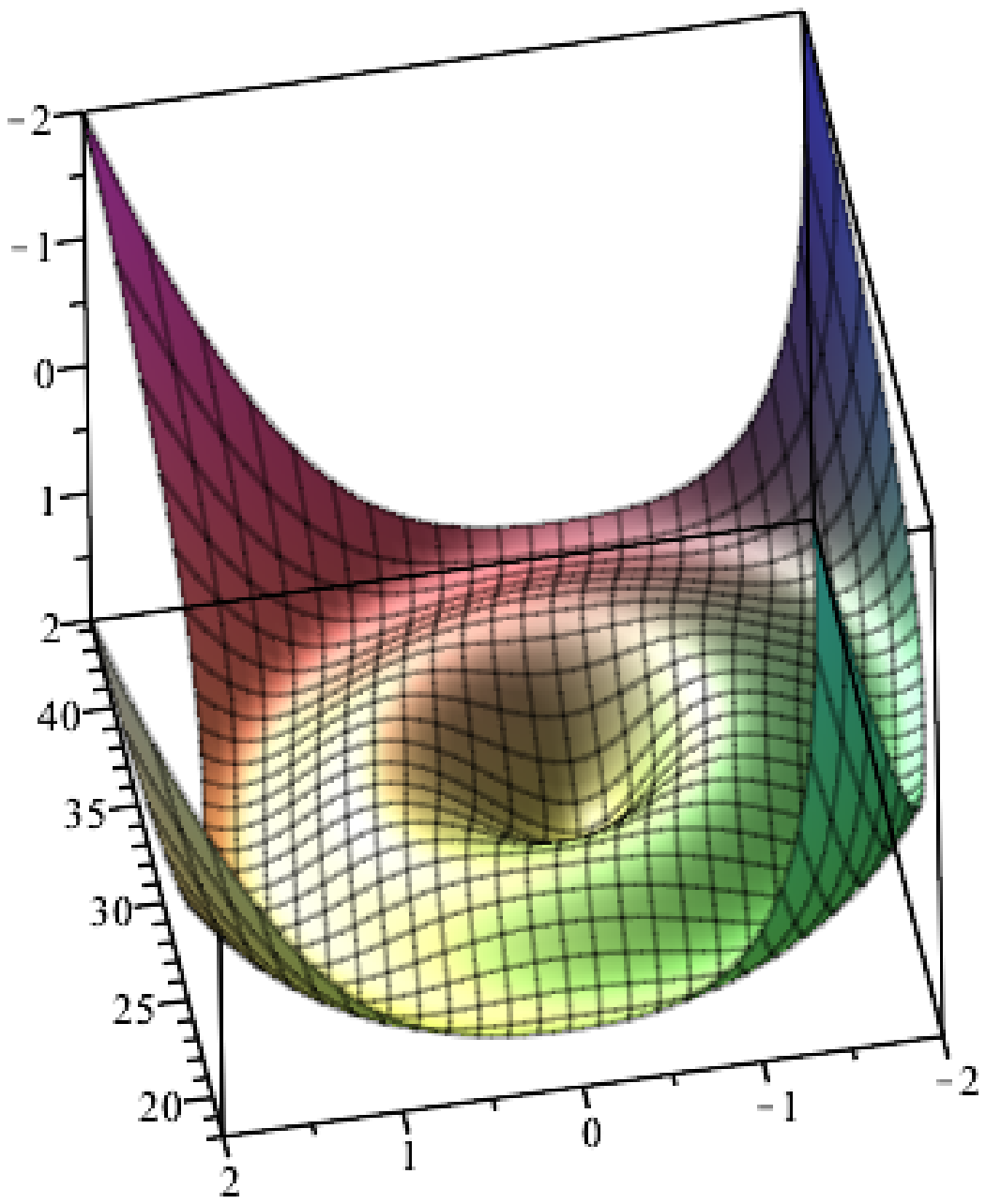}
\includegraphics[width=6cm, height=6cm]{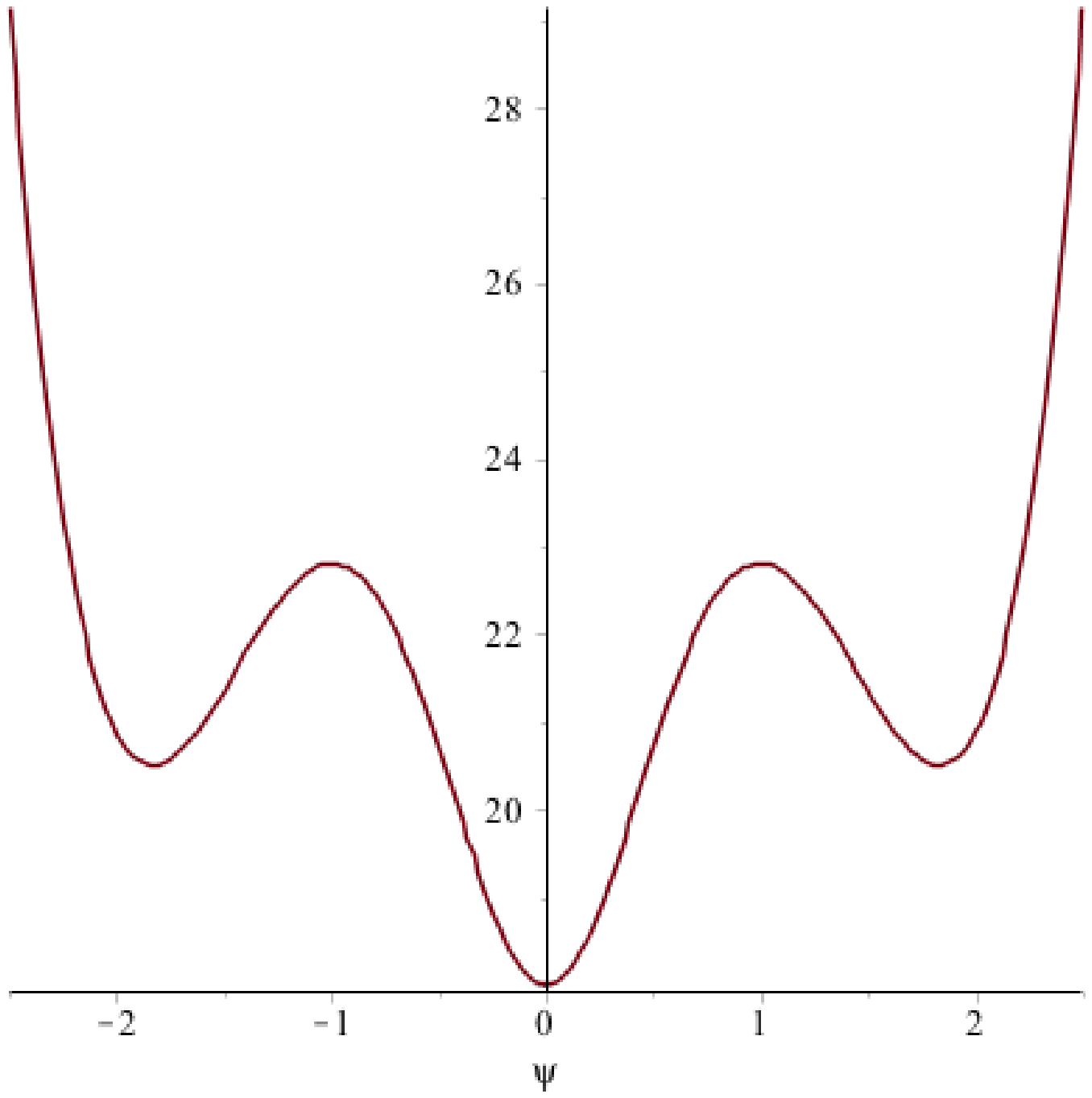}
\caption{
The vacuum value in  $<M>=0.5$,  $\kappa_1 = -8$, $\kappa_2=-6.5$,  $\Delta_{CS}=1 $, $q=1$ and $g=- 0.2$.    }
\end{center}
 \label{Figure3}
 \end{figure}

In Figure 3, we have a configuration where the minimum is isolated and is lower than the minimum circle. In this case, the current is likely to reverse and the electrons migrate to the minimum location. The minimum must be negative, also giving a Hall effect.
 


\section{Chern-Simons fermionic model of graphene-like structures with vortex configuration}

 In this Section, let us consider the fermionic coupling in our model taking into account graphene-like structures. 
We propose a discussion on our supersymmetric model and  graphene-like mass gap. Graphene consists of a sheet composed by atoms with $sp^2$ hybridization so that the crystal is formed by a two-dimensional hexagonal lattice, that can be viewed as the superposition of two sub-networks with a triangular unit cell formed by two carbon atoms, identified as A and B. 
 
 In Table II, we consider the important fermionic field split from (1+3) dimensions. The interesting terms here are the four-component Majorana spinors that can be written in terms of two components Dirac field in (1+2) dimensions.

 (1+2)-dimensional systems have been a strong theoretical and 
experimental interest for the monolayer graphene experimental realization
in 2004 \cite{Novoselov2004,Novoselov2005,Zhang2005},  where  it was 
observed that low-energy excitations behave like negatively-charged 
fermions satisfying a Dirac equation. These new possibilities have opened up a new interest on topological defects in lower-dimensional 
fermionic systems. 

The Dirac-type excitations in pure graphene are gapless, but we can introduce specific impurities that affect the short-distance electron-electron interactions and yield the appearance of massive Dirac fermions \cite{Zhu852012,Araki3022011,Eva2012,Cooper2012}. This illustrates a way to consider topological defects as a mechanism to introduce massive Dirac fermions \cite{Chakraborty2012}.

 The fields with global symmetry may describe graphene-like properties. The Lagrangean we consider takes the form:

 \begin{eqnarray}
 {\cal L}= \Big[ {i \alpha^2 \over 2}(\bar \Psi_+ \gamma^\mu \partial_\mu \Psi_+ + \bar \Psi_- \gamma^\mu \partial_\mu \Psi_-) + {i \over 2} ( \bar \Lambda_+\gamma^\mu \partial_\mu  \Lambda_++ \bar \Lambda_- \gamma^\mu \partial_\mu \Lambda_- ) + V_f^I\Big]
 \end{eqnarray}
 where  $ \alpha^2 = {1\over 2}( 1 - {g^2 \over 2} \phi \phi^* e^{-2 g M})$   and  $V^I_f$ is the fermionic potential given by
 
 \begin{equation}
 V^I _f=  - i m_1 \, \beta \, (\bar \Lambda_+ \Psi_- - \bar  \Lambda_- \Psi_+)-  m_2 (\bar\Psi_+\Psi_+-\bar\Psi_- \Psi_- )
 \end{equation}
 where 
  \begin{eqnarray}
  m_1&=& {  \Delta_{CS}  + g q \phi^* \phi e^{-2g M } \over  \beta}  \\
  m_2 &=& {{g^2 \over 4}\Big[(q  + g \Delta_{CS}) N + {\kappa_2 \over2}\Big] \over (1 - g^2 \phi \phi^* e^{-2gM})} \phi \phi^*e^{-2gM} . 
\end{eqnarray}

 In the case with Dirac fields describing electrons in graphene, we consider 
 \begin{equation}
 \Lambda_{\pm} = \pm \, \beta \, {i \over 2}  \,  \Psi_{\mp} \label{phi}
 \end{equation}

 Finally, one can write  this system as the sum of two Lagrangian
densities, one for each valley:
 
 \begin{eqnarray}
  {\cal L}_+&= &{i v_F \over 2}\bar \Psi_+ \gamma^\mu \partial_\mu \Psi_+ - m \bar \Psi_+ \Psi_+\\
 {\cal L}_-  &=&{ i v_F \over 2}\bar \Psi_- \gamma^\mu \partial_\mu \Psi_- + m \bar \Psi_- \Psi_-
 \end{eqnarray}  
with  $m = m_1 + m_2 $ and  $v_F = \alpha^2 + \beta^2  $.  The two sub-lattices, $\Lambda_A$ and $\Lambda_B$,  for two Dirac points are represented by $ \Psi_+$  and $\Psi_-$. 
The spinor and Pauli matrices that contribute in this case are 
\begin{equation}
\Psi_+ = \left( \begin{array}{ll}
\psi_+^B \\
\psi_+^A
\end{array} \right), \,\,\,\,  \Psi_- = \left( \begin{array}{ll}
\psi_-^A \\
\psi_-^B
\end{array} \right)\label{spinor+-}
\end{equation}
The two flavors are associated to the two Fermi points that constitute the Fermi surface of the neutral material
and, in the standard description, the electron spin is an extra degree of freedom usually disregarded, as long as
the physics in the system does not explicitly depends on it.
We can assembly the Dirac two-component spinors into a four-component Dirac spinor (we are doubling the spinor representation) as
\begin{equation}
 \Psi = \left( \begin{array}{cc}
\psi_+^B  \\
\psi_+^A\\
\psi_-^A\\
\psi_-^B
\end{array} \right) \label{spinortotal}
\end{equation}
We take the $\gamma$-matrix representation as  $\gamma^0 \equiv \sigma_z $, $\gamma^1 \equiv i \sigma_y $, and $ \gamma^2 \equiv  i\sigma _x $,  then we can consider $\gamma^{\mu} = ( \sigma_z, i \sigma _y, i\sigma_x)$  with $ \bar \Psi = \Psi^\dagger \gamma^0$.
\begin{equation}
 \sigma_x = \left( \begin{array}{cc}
0 & 1 \\
1 & 0
\end{array} \right) , \,\,
\sigma_y = \left( \begin{array}{cc}
0 &- i \\
i & 0 
\end{array}\right), \,\,
\sigma_z = \left( \begin{array}{cc}
1 & 0 \\
0 & -1 
\end{array}\right)
\end{equation}

 \begin{figure}[htb]
\begin{center}
\includegraphics[width=6cm, height=6cm]{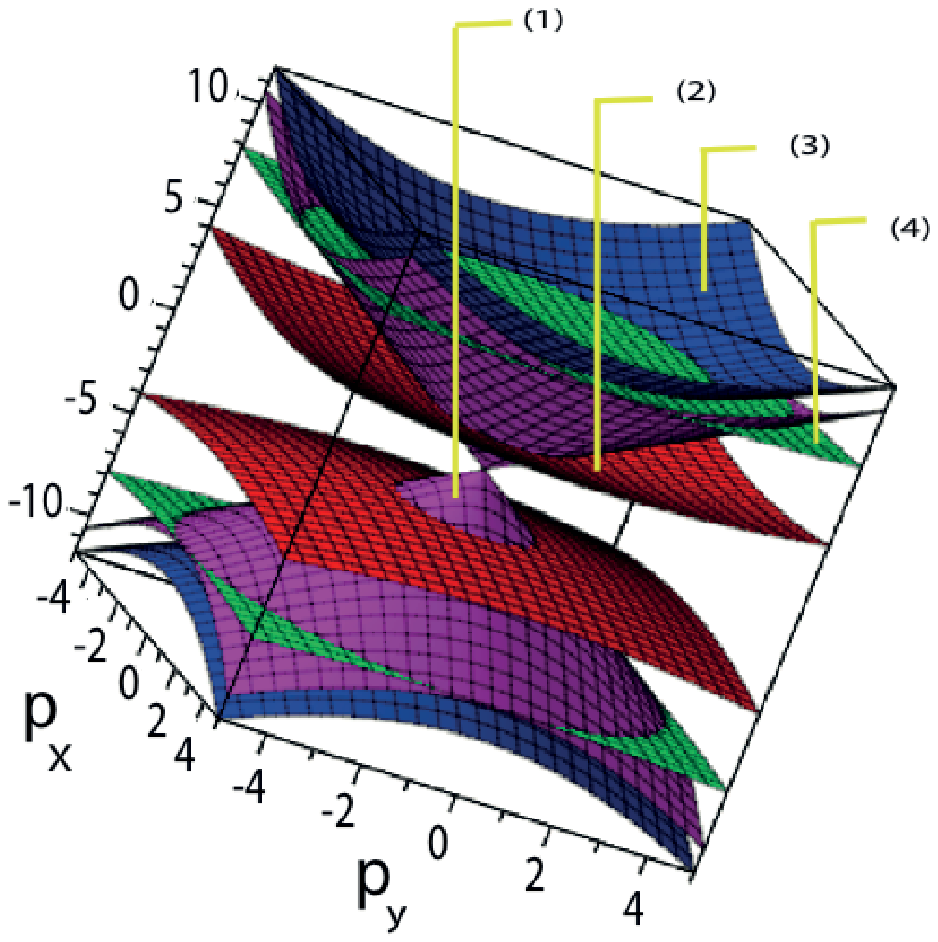}
\includegraphics[width=5cm, height=5cm]{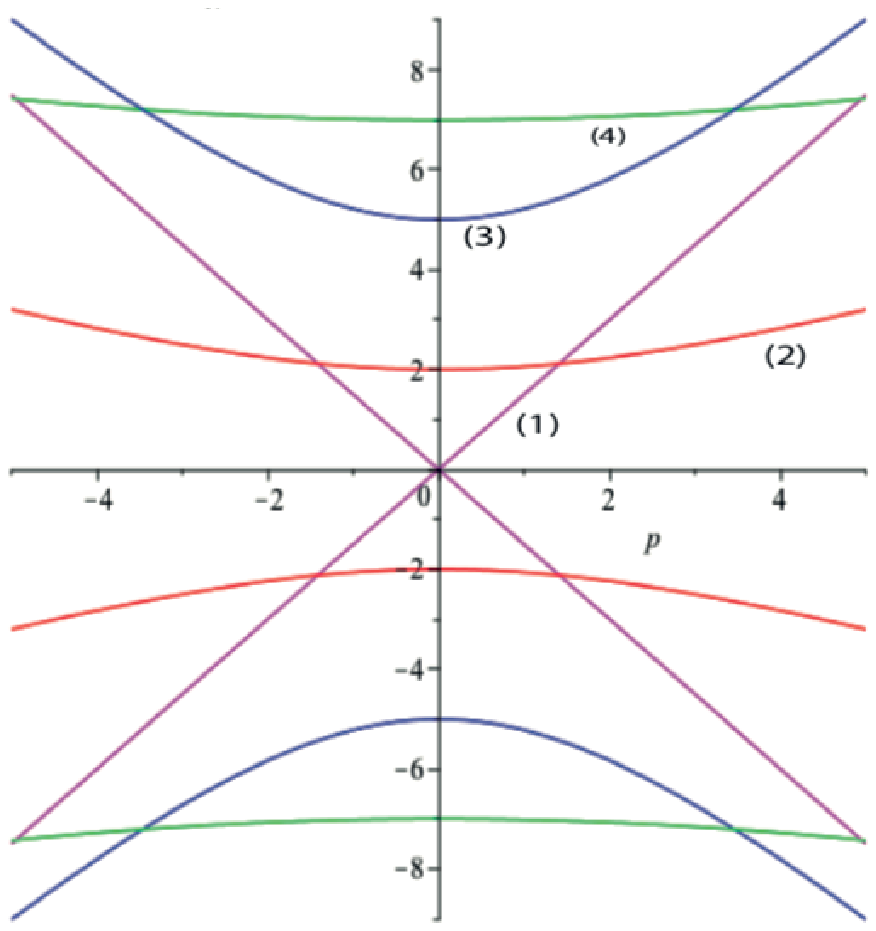}
\caption{Graph for the dispersion equation that represents four-band  type of energy with $q=1$, $\kappa_1=-8$ and $\kappa_2=0$ and M=0, a
for all analyses. The conditions (1) are $\Delta_{CS}=0 $ and $g=0$,  to (2) are $\Delta_{CS}=0 $ and $g =1$ in (3) we have $\Delta_{CS}=10 $ and $g=0$,  to (4) are $\Delta_{CS} =10 $ and $g =1$.}
\end{center}
 \label{Figure 4}
 \end{figure}

 In Figure 4, we depict the case (1), where we have the limit where the vortex is pure, with no Chern-Simons coupling $ \Delta_ {CS} $ contribution and non-minimal coupling contribution, "g". We can see, in this case, that the material behaves as a conductor and there is no energy gap. In case (2), we have vanishing Chern-Simons $ \Delta_ {CS} $ coupling constant, and the coupling constant $ g \neq 0 $. In this case, a mass associated with the constant g shows up. Already in case (3), we can see that, for g = 0, the mass for the system is the Chern-Simons parameter, $ \Delta_ {CS} $. The largest mass corresponds to the case where both $ \Delta_ {CS} $ and $ g $ are nonzero.
 \begin{figure}[htb]
\begin{center}
\includegraphics[width=6cm, height=6cm]{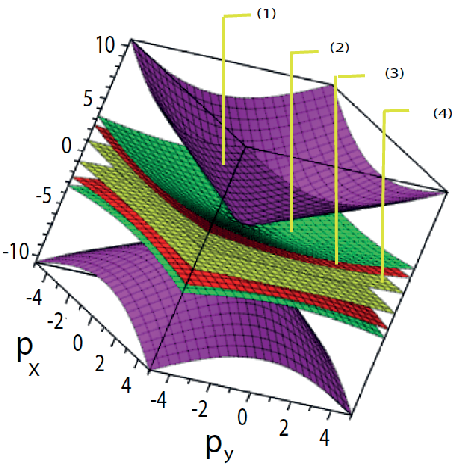}
\includegraphics[width=5cm, height=5cm]{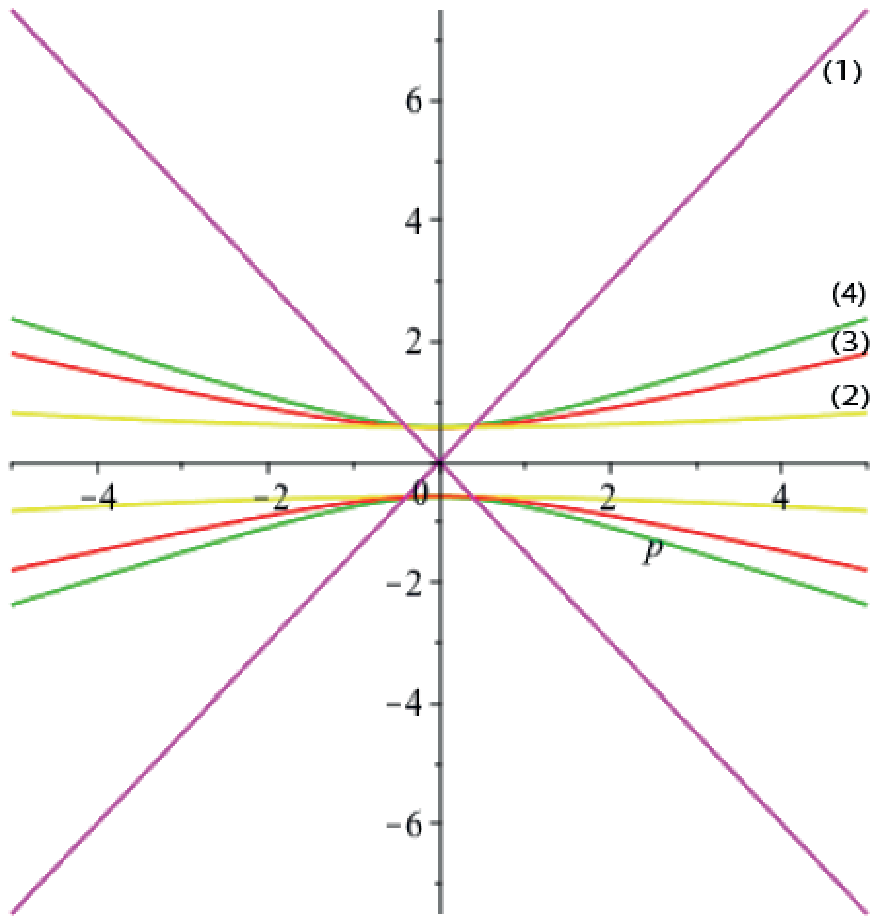}
\caption{Graph for the dispersion equation that represents a four-band  type of energy with $q=1$, $\kappa_1=-8$ and $\kappa_2=-6.5$  and M=1.2 for all analyses. The conditions (1) are $\Delta_{CS}=0$, $g=0$ and $\beta =1$;  for (2), they are $\Delta_{CS} =0.5 $ and $g =-0.2$. In (3), we have $\Delta_{CS}=1 $ and $g=-0.2$;  in (4), they are $\Delta_{CS} =1.2 $ and $g =-0.2$.}
\end{center}
 \label{Figure 5}
 \end{figure}
The case of Figure 5 is the one where the potential presents, beyond the degenerate circle, proper of the dubbed Mexican potential, a minimum in the core of the defect. For the case shown in this Figure, $ M> 1 $, which implies that the central minimum has a larger value of the potential than the degenerate minimum circle. At this point, we have $ \kappa_1 $ and $ \kappa_2 \neq 0 $. In case (1), we have the typical situation where $ \Delta_ {CS} = 0 $ and $ g = 0 $, where the central point is a maximum point and we have the conductive phase. In other cases, we have $ g <0 $ $ \Delta_ {CS} $ assuming several values. As $ \Delta_ {CS} $ decreases, we get smaller line curvatures, with a gap that is pretty much the same.

\section{Spin current in graphene  with a topological insulator phase }
In this Section, we consider the interaction with  a constant external magnetic field.  We discuss the possibility to find a spin current \cite{Cortijo:2010jn} in our treatment. We have a spin current  when there is a flow of the intrinsic spin degrees of freedom.  
The spin current is given by  the difference between the spins that flow up $J_{\uparrow}$ and those flowing down $J_{\downarrow} $. This corresponds  to the Schwinger representation of the spin operator  $  J_s^i = {1 \over 2}  \Psi^\dagger \sigma^i \Psi$.  
When we couple the Dirac fermions to a background field, we can trigger the appearance of a topological insulating phase different from the standard insulator phase. The full Lagrangian that contains the gauge field $A_{\mu}$ is given by

\begin{equation}
{\cal L} = {\cal L}_{gauge} + {\cal L}_{scalar} + {\cal L}_{int}
\end{equation}
where ${\cal L}_{gauge}$ is given by (\ref{CS}) and  ${\cal L}_{scalar}$ is given by  (\ref{scalar}).
We can reproduce the same behavior in our approach with the action

\begin{equation}
{\cal L} _{int}= - { g^2  \over 4q}( \bar \Psi_+  \gamma^{\mu }{\cal J}_{\mu } \Psi_+  -   \bar \Psi_-  \gamma^{\mu }{\cal J}_{\mu } \Psi_- ) e^{-2gM},\label{currentaction},
\end{equation}

\begin{eqnarray}
{\cal J}_\mu & =&-{iq \over 2}\Big(\phi^* \nabla_\mu \phi - \phi (\nabla_\mu \phi)^*\Big) \\
\nabla_{\mu} &=& \partial_\mu + i q A_\mu + i g \epsilon_{\mu \nu \rho} H^{\nu \rho}
\end{eqnarray}
As we already considered in Section 2,  there are two gauge couplings in the action (\ref{currentaction}), as we can observe in (\ref{covderivative}) with the current given by (\ref{current}). One of these appears as the dual of field strength, and it is not a candidate to our external magnetic field. The other one is the usual gauge field, that can be coupled to the fermions by (\ref{currentaction}) and  gives the current. The presence of the $\Gamma_5$ allows the interpretation of this term as the spin-orbit coupling.

The field equation for $A_\mu $ is given by the bosonic part  and the fermionic sector gives us the full equation
\begin{eqnarray}
\partial _{\mu }F^{\mu \nu }= {\cal J}^{\nu } - J_s^\nu + \Delta_{CS}\epsilon^{\nu  \rho \kappa} \partial_{\kappa } B_{\rho},
\end{eqnarray}
with the notation $F_{\mu \nu} $ for the electromagnetic field strength and  $B_{\mu}$ given by

\begin{eqnarray}
F^{\mu \nu} = \left\{ \begin{array}{ll}
F_{0i} \equiv \vec{E}_i\\
F_{i j} \equiv - \epsilon_{ij} B .
\end{array}\right. &\, \, \, \,  B^{\mu} = \left\{ \begin{array}{ll}
B^0 \equiv \chi \\
B^i \equiv \vec{W}_i
\end{array}\right.
\end{eqnarray}

Now, for the Chern-Simons sector, $\epsilon^{\nu  \rho \kappa} \partial_{\kappa } B_{\rho}$  fulfills the following identities

\begin{equation}
\Delta_{CS}\epsilon^{\nu  \rho \kappa} \partial_{\kappa } B_{\rho} = -  \Delta_{CS}\epsilon^{\nu \kappa \rho } \partial_{\kappa } B_{\rho} = -  {\Delta_{CS} \over 2}\epsilon^{\nu \kappa \rho } (\partial_{\kappa } B_{\rho} - \partial_{\rho } B_{\kappa}) = -  {\Delta_{CS} \over 2}\epsilon^{\nu \kappa \rho }  H_{\kappa \rho}
\end{equation}
with $H_{\kappa \rho}= \partial_{\kappa } B_{\rho} - \partial_{\rho } B_{\kappa}$ given by

\begin{eqnarray}
H_{\kappa \rho} = \left\{ \begin{array}{ll}
 H_{0i} =  \vec{ e} = - \vec{\nabla} \chi - \partial_t \vec{W}\\
 H_{i j} = - \epsilon_{ij } b = - \epsilon_{i j} \vec{\nabla } \times \vec{ W}
 \end{array} \right.
 \end{eqnarray}

$J^\mu$ is the current given by the charged fermions and  bosons and $J_S^\mu $ reads as shown below: 
\begin{eqnarray}
J_S^\mu =  a \Big( \bar \Psi_+ \gamma^\mu \Psi_+ -  \bar \Psi_- \gamma^\mu \Psi_- \Big) = J^\mu_{\uparrow} - J_{\downarrow}^{\mu}
\end{eqnarray}
with $ a=   {q g^2 \phi^*\phi \over4} e^{-2g M}$, when we open in components considering that the component $J^0_S$ is the fermionic density and $J^i_S$ is the fermionic vector-like current, then we have

\begin{eqnarray}
\partial _{i}F^{i  0}= {\cal J}^{0 } - J_S^0 + \Delta_{CS}\, \epsilon^{0  i j} \partial_{i} B_{j}\\
\vec{\nabla}  \cdot  \vec{ E}_{ef} = \rho  - \rho_S
\end{eqnarray}
with $  \vec{E} _{ef} = \vec{E}  - \Delta_{CS }  \tilde{\vec{W}}$, with $\rho $ is related with the electric charge and $\rho_S$ is the spin density given by
\begin{eqnarray}
\rho_S =  a\, \Big( \bar \Psi_+ \sigma_z \Psi_+ -  \bar \Psi_- \sigma_z \Psi_- \Big) \\
\rho = -{i q  \over 4} [\phi^*\nabla_0 \phi - \phi (\nabla_0 \phi)^*] \label{eletrccurrent}
\end{eqnarray}

We can see that the Gauss's law in the presence of the Chern-Simons term presents an effective electric field with a typical polarization vector term determined by the dual vector $ \tilde {\vec {W}} $. We then have two types of densities, one of them due to the couplings to the defect fields which, in the case of an uncharged defect, disappears. We have the spin density current. Thus, there is a spin bias whose total density is nonzero. In this case, we have that there is an unbalance of spins in the system, which can create magnetization.

The other possibility corresponds to the case where we have the dynamical magnetic field in the presence of an external electric field; in this case, we have 
\begin{eqnarray}
\partial _{\mu}F^{\mu  i}= {\cal J}^{i } - J_S^i - {\Delta_{CS} \over 2}    \, \epsilon^{i \kappa \rho}  H_{\kappa \rho}\\
\tilde \partial_i B_{ef}  = \partial_t \vec{E}_{ef}+ \vec{{\cal J}}  - \vec{J}_S  \label{maxwellcurrent}
\end{eqnarray}
with $ B_{ef} = B + \Delta_{_{CS}} \chi$, $\tilde \partial_i = \epsilon_{ij} \partial_j $.

The equivalent Maxwell's equation in (1+2) dimensions is given by the equation (\ref {maxwellcurrent}), which exhibit two currents. One of them is associated to the coupling with the defect fields, the other one is nothing but a spin current due to the unbalanced spins.

\section{Concluding  Comments}

The main purpose of this work is the investigation of possible solutions to a (1+2)-dimensional Chern-Symons model in connection with non-minimal coupling. The idea is to carry out the dimensional reduction of a supersymmetric (1+3)-dimensional model containing a spinor superfield, dubbed Kalb-Ramond supermultiplet. In this model, we have some important features that influence the solution. 
Under certain conditions, we show that it is possible to find a potential that exhibits two types of minima. One of them is the usual circle of minima of the topological defect, which is the Mexican hat-type. An new feature of the model is the appearance of an absolute minimum in the core of the defect. We have experimental evidence of the emergence of a current on the edges of a graphene-like material whenever it is subject to conical deformations. In this case, the material behaves as a topological insulator with surface currents. In the literature, as mentioned in the Introduction, the material behaves like in the case of a vortex in presence of a gravitational field with a deficit angle that is related to the surface current \cite{Takeuchi:2012ee,Fonseca:2011fe}. 
 A new result of this work is the proposal of a theoretical model for the symmetry breakings and their relation to the vortex formation and the effects on the charge and spin currents. From an exclusively gravitational point of view, these breakings are not possible to be understood. The presence of a potential with this pattern of minima is responsible for  the appearance of two types of polarization verifiers. One of them occurs whenever the minimum of the potential in the core is greater than the minimum circle; in this case, the polarization vector is from top to bottom. The other case occurs whenever the absolute minimum is less than the circle of minimums; in this case, the polarization vector is inverted.
 We have also analyzed the energy bands of the system related to the  parameters present in the potential of the topological defect - a vortex. We can see, in Figures 2 and 3, that the M-field, coming from the Kalb-Ramond sector of the 4-dimensional supermultiplet,  gives us the possibility of having the Hall current. In the fermionic part, it is possible to understand, for the same configurations shown in the potential, that the energy bands decrease to $M \neq 0$. Thus, we see that the M-field can act as a control parameter, as in a semiconductor. This can control the current that appears on the surface of the material.
 Finally, we have studied the types of current that may appear in this category of models. In this case, we focus on the fermionic sector of the model and the couplings of the fermions to currents . We have found that two types of currents may be identified: one due to moving loads and another one due to spin. Thus, we see that the appearance of spin and charge currents is due to the interaction terms.
We hope that these new results can shed some light on graphene-like materials; by analogy, we hope that this sort of model may help in understanding some highly energetic events that may take place in the Universe. As a near-future follow-up of this paper, the idea is to go deeper into the applicability of the theory in these two contexts, that is, in condensed matter physics and the study of gravitational phenomena.

\textbf{Acknowledgements:}

C. N. F.  would like to thank the CBPF for the hospitality during this work and IFF for providing the infrastructure for its completion.

\end{document}